\documentclass[%
reprint,
superscriptaddress,
showpacs,
preprintnumbers,
 bibnotes,
 amsmath,amssymb,
 aps,
prl
]{revtex4-1}

\usepackage{graphicx}
\usepackage{dcolumn}
\usepackage{bm}
\usepackage{color}

\begin{document}


\title{Longitudinal spin excitations and magnetic anisotropy in antiferromagnetically ordered BaFe$_2$As$_2$
}

\author{Chong~Wang}
\affiliation{International Center for Quantum Materials, School of Physics, Peking University, Beijing 100871, China}
\author{Rui~Zhang}
\affiliation{Beijing National Laboratory for Condensed Matter Physics, Institute of Physics, Chinese Academy of Sciences, Beijing 100190, China}
\author{Fa~Wang}
\affiliation{International Center for Quantum Materials, School of Physics, Peking University, Beijing 100871, China}
\author{Huiqian~Luo}
\affiliation{Beijing National Laboratory for Condensed Matter Physics, Institute of Physics, Chinese Academy of Sciences, Beijing 100190, China}
\author{L.~P.~Regnault}
\affiliation{SPSMS-MDN, UMR-E CEA/UJF-Grenoble 1, INAC, Grenoble F-38054, France}
\author{Pengcheng Dai}
\email[]{pdai@rice.edu}
\affiliation{Department of Physics and Astronomy, Rice University, Houston, Texas 77005, USA}
\affiliation{Beijing National Laboratory for Condensed Matter Physics, Institute of Physics, Chinese Academy of Sciences, Beijing 100190, China}
\author{Yuan~Li}
\email[]{yuan.li@pku.edu.cn}
\affiliation{International Center for Quantum Materials, School of Physics, Peking University, Beijing 100871, China}


\begin{abstract}
We report on a spin-polarized inelastic neutron scattering study of spin waves in the antiferromagnetically ordered state of BaFe$_2$As$_2$. Three distinct excitation components are identified, with spins fluctuating along the $c$-axis, perpendicular to the ordering direction in the $ab$-plane, and parallel to the ordering direction. While the first two ``transverse'' components can be described by a linear spin-wave theory with magnetic anisotropy and inter-layer coupling, the third ``longitudinal'' component is generically incompatible with the local moment picture. It points towards a contribution of itinerant electrons to the magnetism already in the parent compound of this family of Fe-based superconductors.
\end{abstract}


\maketitle

\begin{figure}
\includegraphics[width=3.375in]{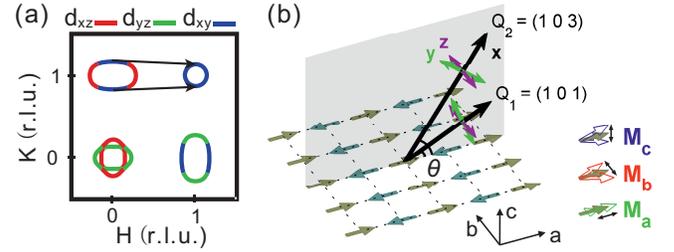}
\caption{\label{fig:zero}
(a) Fermi surface of BaFe$_2$As$_2$, reproduced from \cite{DaiNPhys2012} using band structure from \cite{GraserPRB2010}. Arrows indicate nesting vectors. (b) Spin arrangement and fluctuation directions in the AF phase of BaFe$_2$As$_2$. Coordinate systems for neutron polarization are indicated for two examples $\mathbf{Q}_1$ and $\mathbf{Q}_2$.
}
\end{figure}

Among very different classes of materials including the Fe-based superconductors (FeSC), the cuprates, and the heavy-Fermion compounds, a striking feature of unconventional superconductivity is that it commonly appears close to an antiferromagnetic (AF) phase \cite{Uemura_NatureMater_2009}.
Since magnetism may be a common thread for the pairing interaction in unconventional superconductors \cite{ScalapinoRMP2012}, it is important to determine the microscopic origin of the AF order. For the cuprates, it is well accepted that their Mott insulating parent compounds have localized moments, and the spin waves can be well described by a Heisenberg model \cite{LeeRMP2006,ColdeaPRL2001,HeadingsPRL2010}.  In the case of iron pnictide families of FeSC, there is no consensus on the origin of the stripe-like AF order in the parent compounds \cite{Huang_PRL_2008,Cruz_Nature_2008,ZhaoNPhys2009,Luetkens_NatureMater_2009}. On the one hand, these are semi-metals with hole- and electron-like Fermi pockets at the Brillouin zone center and zone corners, respectively (Fig.~\ref{fig:zero}a) \cite{HirschfeldRepProgPhys2011,LiuPRL2008,YangPRL2009,Liu_PRB_2009}, and the AF order (Fig.~\ref{fig:zero}b) may arise from nesting between the pockets \cite{HirschfeldRepProgPhys2011}, much like the spin-density-wave (SDW) order in chromium \cite{FawcettRMP1988}. On the other hand, the bad-metal phenomenology of iron pnictides \cite{BasovNPhys2011} suggests that these materials are near a Mott transition with magnetism arising from localized moments, much like in the cuprates \cite{Si_PRL_2008,seo_PRL_2008,SachdevJPCM2012}.

If the AF order in the iron pnictides arises entirely from localized moments on Fe, spin waves from these moments should be purely transverse spin excitations (TSE), with moments fluctuating perpendicular to the staggered magnetization keeping an unchanged magnitude.  In contrast, if Fermi surface nesting and itinerant electrons contribute significantly to the AF order, one would expect the presence of longitudinal spin excitations (LSE) with fluctuating moment sizes \cite{SokoloffPR1969a,*SokoloffPR1969b,Knolle_PRB_2010,Kaneshita_PRB_2010,You_PRB_2011}, similar to the LSE seen in the SDW state of chromium \cite{BurkePRL1983,*BoniPRB1998}. Although unpolarized inelastic neutron scattering (INS) experiments have mapped out spin waves in the iron-pnictides parent compounds CaFe$_2$As$_2$ \cite{DialloPRL2009,ZhaoNPhys2009}, BaFe$_2$As$_2$ \cite{Harriger_PRB_2011}, and SrFe$_2$As$_2$ \cite{EwingsPRB2011}, the spectra can be described  by either local-moment \cite{ZhaoNPhys2009,Harriger_PRB_2011} or itinerant models \cite{DialloPRL2009,EwingsPRB2011,Kaneshita_PRB_2010}. To conclusively determine if itinerant electrons contribute to the magnetism, one needs to perform spin-polarized INS experiments to search for LSE in the AF ordered state.  In spite of considerable efforts in this direction on BaFe$_2$As$_2$ \cite{Qureshi_PRB_2012} and NaFeAs \cite{SongUnpublished}, experimental detection of LSE has remained inconclusive so far.

Here we present a spin-polarized INS study of BaFe$_2$As$_2$ in the AF phase, where the ordered moments are aligned along the $a$-axis direction of the orthorhombic structure (Fig. 1b). By comparing magnetic signals that consist of different projections of the intrinsic response, we identify three distinct spin-excitation components with magnetic moments fluctuating along the three crystallographic axes, $M_a$, $M_b$, and $M_c$ (Fig.~\ref{fig:zero}b). The latter two TSE components can be quantitatively described by a linear spin-wave model with magnetic anisotropy. The presence of the (hitherto undetected) LSE component $M_a$, which amounts to about $10\%$ of the low-energy spectral weight, indicates a clear contribution from itinerant electrons. Therefore, itinerant electrons important for superconductivity also contributes to the magnetism in the parent compounds of iron pnictides.

A total of 18 grams of high-quality BaFe$_2$As$_2$ single crystals were grown by a self-flux technique \cite{ChenSupercondSciTech2011} and coaligned with reciprocal lattice vectors $(H~0~L)\equiv H \mathbf{a^*} + L\mathbf{c^*}$ in the horizontal scattering plane. Here we use the orthorhombic crystallographic notation, in which the two-dimensional AF wave vector ($\mathbf{Q}_\mathrm{AF}$) corresponds to $|H|=1$, and the AF zone center and boundary along $\mathbf{c^*}$ correspond to odd and even integer $L$ values, respectively. Our sample has an AF ordering temperature ($T_\mathrm{N}$) of about 137 K and a mosaic of about $1.2^\circ$ (Fig.~S1 in \cite{Supplemental}). The INS experiment was performed on the triple-axis spectrometer IN22 at the Institut Laue-Langevin, France. Heusler crystals were used as spin-polarizing monochromator and analyzer, and CryoPAD was used for performing longitudinal polarization analysis. A flipping ratio of about 16 was maintained throughout our experiment. All measurements were performed in the spin-flip (SF) geometry at a temperature of 2 K. In the AF ordered phase, BaFe$_2$As$_2$ forms randomly distributed orthorhombic twin domains rotated 90$^\circ$ apart. We are effectively not sensitive to half of the sample that develops AF order at $(0~\pm1~1)$ since the spin waves at $(1~0~L)$ are well above the energy range of our measurement \cite{Harriger_PRB_2011}.

In the conventional coordinate system for the neutron spin polarization ($\mathbf{S}$), $\hat{x}$ is along the momentum transfer ($\mathbf{Q}$), $\hat{z}$ is vertical, and $\hat{y}$ is perpendicular to both $\hat{x}$ and $\hat{z}$ (Fig.~\ref{fig:zero}b). Since SF scattering probes magnetic fluctuations perpendicular to both $\mathbf{Q}$ and $\mathbf{S}$, signals that correspond to fluctuations projected along $\hat{y}$ ($\sigma_y$) and $\hat{z}$ ($\sigma_z$) can be obtained by two independent methods: $\sigma_y=\mathrm{SF}_z-\mathrm{BG}=\mathrm{SF}_x-\mathrm{SF}_y$ and $\sigma_z=\mathrm{SF}_y-\mathrm{BG}=\mathrm{SF}_x-\mathrm{SF}_z$, where $\mathrm{SF}_\alpha$ denotes SF intensity measured with incident neutron spins along the $\alpha$ direction, and BG is background intensity. Both methods give consistent results in our study. $\sigma_y$ and $\sigma_z$ are related to the intrinsic magnetic response via
\begin{equation}\label{eqn:one}
\sigma_z=M_b,\,\, \sigma_y=M_{c}\cos^2\theta +M_a\sin^2\theta,
\end{equation}
where $\theta$ is the angle between $\mathbf{Q}$ and $\mathbf{a^*}$ \cite{LuoPRL2013}.

\begin{figure}
\includegraphics[width=3.375in]{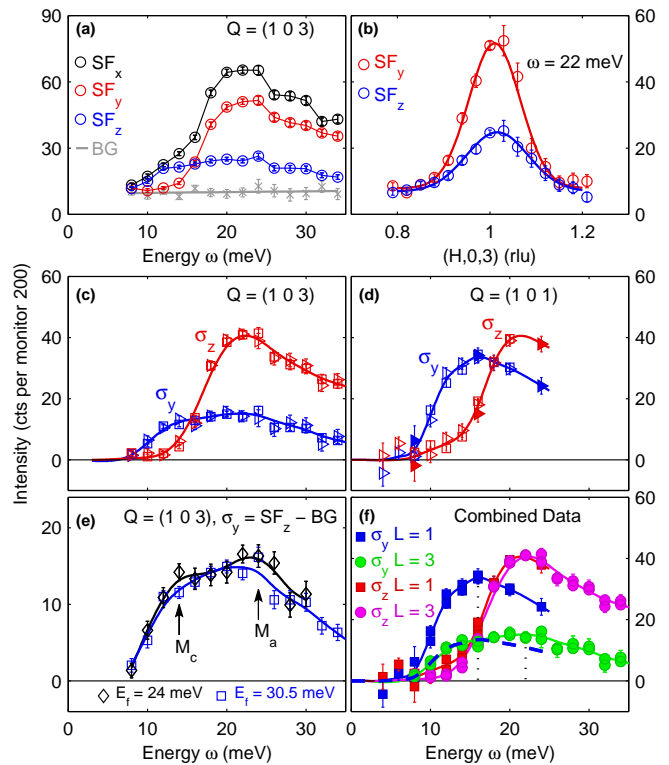}
\caption{\label{fig:one}
(a) Energy scans at $(1~0~3)$. BG is determined by fitting the $\mathrm{SF}_y+\mathrm{SF}_z-\mathrm{SF}_x$ intensity, which does not contain any magnetic signal, to a linear function of $\omega$. (b) $\mathbf{Q}$ scans at $\omega=22$ meV fitted to a single Gaussian peak.  (c-d) Extracted $\sigma_y$ and $\sigma_z$ at $(1~0~3)$ and $(1~0~1)$. Meanings of symbols in (a-d): circles are raw data, squares are $\mathrm{SF}_y$ or $\mathrm{SF}_z$ minus BG, triangles are $\mathrm{SF}_x$ minus $\mathrm{SF}_y$ or $\mathrm{SF}_z$, empty and filled symbols are measured with final neutron energies $E_\mathrm{f}=30.5$ and 50 meV, respectively. (e) Comparison of $\sigma_y$ data obtained with $E_f=24$ and 30.5 meV. (f) Combined data from (c) and (d). Dashed line indicates $\sigma_y$ at $(1~0~3)$ after multiplying by a factor of 0.4. Dotted lines indicate maxima of $\sigma_y$. Solid lines in (c-f) are guide to the eye.
}
\end{figure}

Figure \ref{fig:one}a-b displays raw data of energy and momentum scans at the AF zone center with $L=3$. The extracted $\sigma_y$ and $\sigma_z$ (Fig.~\ref{fig:one}c) exhibit different energy gaps, consistent with an earlier report \cite{Qureshi_PRB_2012}. The result of similar measurement and analysis at $L=1$ is shown in Fig.~\ref{fig:one}d. A quantitative comparison between these measurements is presented in Fig.~\ref{fig:one}f, where the different $\theta=23.5^\circ$ and $52.5^\circ$ (for $L=1$ and 3, respectively) determines the amount of $M_a$ and $M_c$ contributions to $\sigma_y$ (Eq.~(\ref{eqn:one})). The excellent agreement between the $\sigma_z$ data is consistent with a negligible variation in the magnetic form factor (Fig.~S2 in \cite{Supplemental}) and in the instrument resolution from $L=1$ to $L=3$.  A clear difference is found between the $\sigma_y$ data apart from the overall intensity change: At $L=1$, $\sigma_y$ exhibits a maximum at 16~meV, above which the signal decreases in a fashion similar to the decrease of $\sigma_z$ above 22~meV. At $L=3$, while $\sigma_y$ exhibits a rapid increase between 8 and 14~meV similar to the behavior at $L=1$, it continues with a ``plateau-like'' profile to higher energies, and reaches a global maximum at around 22~meV.  If $\sigma_y$ consists of only $M_c$, the data for $L=1$ and $L=3$ are expected to be identical after multiplying the former by a factor of 0.44, which accounts for the difference in $\theta$. We find the best agreement between the two data sets below 16~meV by multiplying the $L=1$ data by 0.40. The normalized $\sigma_y$ at $L=1$ (dashed line in Fig.~\ref{fig:one}f) lies below the $L=3$ data above 18~meV. We attribute this difference to a non-zero contribution from $M_a$, which, unlike $M_c$, is expected to increase by a factor of 4 from $L=1$ to $L=3$. To further verify this interpretation, we measured at $(1~0~3)$ with a smaller $E_\mathrm{f}=24$~meV (Fig.~\ref{fig:one}e). Indeed, the improved energy resolution ($\sim3.1$~meV at $\omega=20$ meV, compared to $\sim3.9$~meV for $E_\mathrm{f}=30.5$~meV) leads to a clearer separation of the $M_c$ and $M_a$ components. These results establish the presence of an $M_a$ contribution to the total magnetic response at the AF zone center.

\begin{figure}
\includegraphics[width=3.175in]{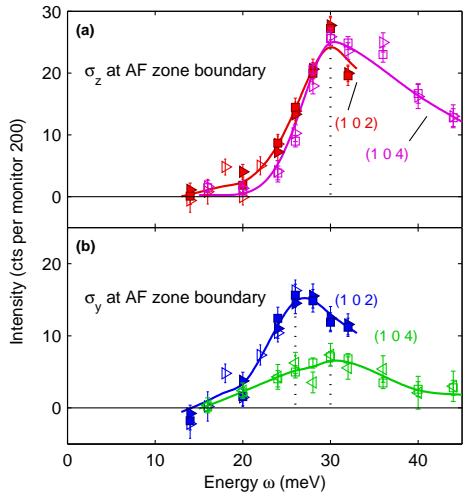}
\caption{\label{fig:two}
$\sigma_z$ (a) and $\sigma_y$ (b) at AF zone boundaries (see Fig.~S3 in \cite{Supplemental} for the raw data). The meanings of symbols are the same as in Fig.~\ref{fig:one}. Solid lines are guide to the eye.
}
\end{figure}

BaFe$_2$As$_2$ consists of FeAs layers separated by Ba. The magnetic coupling $J_c$ between neighboring layers gives rise to a spin-wave dispersion along $\mathbf{c^*}$, with a saddle point at the AF zone boundary where $J_c$ can be best determined. In a recent unpolarized INS measurement at the AF zone boundary \cite{Park_PRB_2012}, a substantially smaller $J_c$ was found than previously inferred \cite{ZhaoNPhys2009,Harriger_PRB_2011}. Spin-polarized measurements have not been attempted at the AF zone boundary since this refined study. In addition to the search for LSE, such measurements provide a stringent test of spin-wave models for describing the TSE, with additional constraints on the model parameters.

Figure~\ref{fig:two} displays the extracted $\sigma_y$ and $\sigma_z$ at the AF zone boundaries. For $L=2$, measurements with different $E_\mathrm{f}$ are combined to satisfy the scattering kinematic constraint for the extended energy range. Additional tests (Fig.~S3 in \cite{Supplemental}) show no evidence for a distortion of data due to instrument resolution. The results are qualitatively similar to those at the AF zone center: (1) At $L=2$, the maximum of $\sigma_z$ occurs at a higher energy than $\sigma_y$, and both energies are higher than the corresponding values at $L=1$. (2) $\sigma_z$ are nearly identical at $L=2$ and $L=4$. (3) $\sigma_y$ reaches its maximum at a higher energy at $L=4$ than at $L=2$. From $L=2$ to $L=4$, one expects a decrease by $56\%$ in the contribution of $M_c$, and an increase by $74\%$ in the contribution of $M_a$. The data in Fig.~\ref{fig:two}b are thus consistent with maxima of $M_c$ and $M_a$ contributions at around 26 and 30~meV, respectively.

\begin{figure}
\includegraphics[width=3.175in]{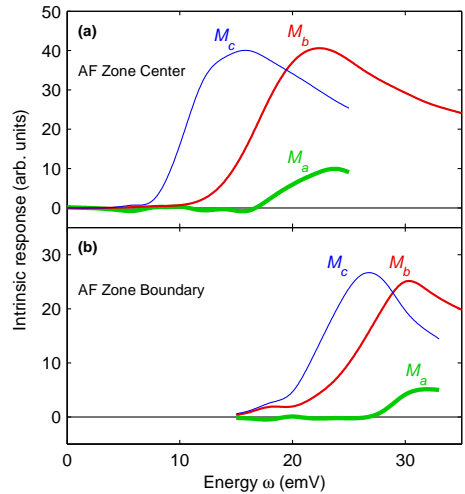}
\caption{\label{fig:three}
Intrinsic magnetic responses (see Fig.~1 for definition) calculated from the interpolated data in Figs.~2 and 3.
}
\end{figure}

To start a quantitative discussion, we first plot in Fig.~\ref{fig:three} the intrinsic magnetic responses at the AF zone center and boundary: $M_b$ is determined from the average of the interpolated $\sigma_z$ data at $L=1$, 3 and 2, 4 in Figs.~2 and 3, respectively. $M_a$ and $M_c$ are calculated from the interpolated $\sigma_y$ data using Eq.~(\ref{eqn:one}), which allows us to find a unique solution given two measurements with different $\theta$. The energies of the TSE spin waves ($E_\mathrm{exp}$, the energy where the signal reaches $90\%$ of the maximum \cite{Qureshi_PRB_2012}) are identified from the $M_c$ and $M_b$ data and summarized in Table.~\ref{table:one}. In terms of spectral weight, $M_c$ and $M_b$ are roughly equal, and their values at the AF zone boundary are about $30\%$ smaller than at the zone center. The latter observation is expected because in linear spin-wave theory the intensity of low-energy excitations is inversely proportional to the energy. However, as has been pointed out by Qureshi \textit{et al.} \cite{Qureshi_PRB_2012}, the equal amplitudes of $M_c$ and $M_b$, despite their energy difference, are inconsistent with the linear spin-wave theory, and might indicate a necessity of resorting to more sophisticated calculations that also include itinerant electrons. Indeed, a clear LSE $M_a$ component is found at both the AF zone center and the zone boundary at roughly the same energies as $M_b$, and it amounts to about $10\%$ of the low-energy spectral weight. To our knowledge, this is the first direct evidence for a contribution from itinerant electrons to the spin excitations in a FeSC parent compound \footnote{A measurement performed by Qureshi \textit{et al.} \cite{Qureshi_PRB_2012} at $(1~1~5)$ is (in hindsight) consistent with an $M_a$ contribution at the AF zone center (Fig.~S4 in \cite{Supplemental}). The low signal-to-noise ratio had prevented the authors from identifying the effect.}.

\begin{table}[tbp]
 \begin{tabular}{|c|c|c|}\hline\hline
~~$M$~~        &$E/2S$ &$E_\mathrm{exp}\,$/$\,E_\mathrm{cal}$ \\\hline
$M_{b,L=1}$ 		 &$\sqrt{(D_{x}-D_{y})(4J_2+2J_{1a}+2J_c+D_{x})}$	&18.9\,/\,18.8\\\hline
$M_{c,L=1}$   	    &$\sqrt{D_{x}(4J_2+2J_{1a}+2J_c+D_{x}-D_{y})}$	&11.6\,/\,11.7\\\hline
$M_{b,L=2}$ 	   	 &$\sqrt{(D_{x}-D_{y}+2J_c)(4J_2+2J_{1a}+D_{x})}$	&28.3\,/\,28.5\\\hline
$M_{c,L=2}$         &$\sqrt{(D_{x}+2J_c)(4J_2+2J_{1a}+D_{x}-D_{y})}$	&24.6\,/\,24.4\\\hline
\hline
\end{tabular}
\caption{\label{table:one}
Spin-wave energies calculated from Eq.~(\ref{eqn:two}), using $SJ_{1a}=59.2$ meV, $SJ_2=13.6$ meV, $SJ_c=0.333$ meV, $SD_x=0.196$ meV, and $SD_y=-0.311$ meV. The last column shows comparison with experimental data (in meV). Derivation of the expressions is given in \cite{Supplemental}.
}
\end{table}

We consider the following Heisenberg Hamiltonian \cite{Supplemental} for a quantitative description of the TSE:
\begin{equation}\label{eqn:two}
\begin{split}
&H=\sum_{\mathbf{r}}[J_{1a}\mathbf{S}_{\mathbf{r}}\cdot\mathbf{S}_{\mathbf{r}+\hat{x}}+J_{1b}\mathbf{S}_{\mathbf{r}}\cdot\mathbf{S}_{\mathbf{r}+\hat{y}}+
J_{2}(\mathbf{S}_{\mathbf{r}}\cdot\mathbf{S}_{\mathbf{r}+\hat{x}+\hat{y}}+\\
&\mathbf{S}_{\mathbf{r}}\cdot\mathbf{S}_{\mathbf{r}-\hat{x}+\hat{y}})-D_{x}{(S_{\mathbf{r}}^{x})}^2-D_{y}{(S_{\mathbf{r}}^{y})}^2+J_{c}\mathbf{S}_{\mathbf{r}}\cdot\mathbf{S}_{\mathbf{r}+\hat{z}}]
\end{split}
\end{equation}
where $J_{1a}$ and $J_{1b}$ are the nearest-neighbor interaction along the $a$ and $b$ directions, respectively, $J_2$ is the next-nearest-neighbor interaction, and $J_c$ is the inter-layer coupling. $D_{x}$ and $D_{y}$ denote the single-ion anisotropy. In Table~\ref{table:one} we list the expressions for the spin-wave energies. The exchange coupling parameters, $J_{1a}$, $J_{1b}$, and $J_2$, have been determined from time-of-flight INS measurement \cite{Harriger_PRB_2011}. Our data allow us to determine the remaining three parameters with four constraints. The fitted $SJ_c$ value  (Table.~\ref{table:one}) is consistent with the report by Park \textit{et al.} \cite{Park_PRB_2012}, taking into account the slightly different criteria of defining the spin-wave energies, and $SD_x$ and $SD_y$ are consistent with the report of Qureshi \textit{et al.} \cite{Qureshi_PRB_2012}. Our experimental result can also be described by exchange anisotropy instead of single-ion anisotropy \cite{Supplemental}, but since the two types of anisotropy give nearly identical spin-wave dispersions, they cannot be distinguished by INS measurements.

To understand the physical origin of the LSE, we first note that the AF order in BaFe$_2$As$_2$ is commensurate. This is different from the incommensurate SDW order in chromium, and it precludes an interpretation of the LSE as phason modes \cite{FishmanPRB1996}. Another possible form of LSE in itinerant antiferromagnets is the amplitude mode \cite{SokoloffPR1969a,*SokoloffPR1969b}. The lowest energy required to create such excitations occurs at the AF wave vector, consistent with our finding, and it is equal to twice the energy gap ($2\Delta_\mathrm{SDW}$) between the magnetically split bands \cite{SokoloffPR1969a,*SokoloffPR1969b,FishmanPRB1994,Knolle_PRB_2010,Kaneshita_PRB_2010,You_PRB_2011}. In the AF ordered phase, optical measurements reveal $\mathbf{q}=0$ interband transitions at 45 and 110~meV which possibly indicate gap opening \cite{Hu_PRL_2008,ChenPRL2010}, but photoemission studies show a complicated band reorganization \cite{YangPRL2009,Liu_PRB_2009,RichardPRL2010} without clear gap opening at the Fermi level \cite{Liu_PRB_2009}. Since the excitations relevant to our INS data occur at finite $\mathbf{q}=\mathbf{Q}_\mathrm{AF}$ with $L=0$ or 1, the correspondence between our data and $2\Delta_\mathrm{SDW}$ values inferred from optical measurements is not obvious, especially since the band structure exhibits a finite $k_z$ dependence \cite{LiuPRL2009}. The similar energies of $M_a$ and $M_b$ (Fig.~\ref{fig:three}) implies a connection between the energy scales of the itinerant and the localized electron systems. The fact that we do not observe a clear decrease of $M_a$ up to the highest energy of our measurements (Fig.~\ref{fig:three}) is consistent with the expectation that the observed $M_a$ is at the bottom of an LSE continuum \cite{FishmanPRB1994,Knolle_PRB_2010,You_PRB_2011}. Finally, we note that the energy of $M_a$ at the AF zone center is consistent with a transient optical response frequency at the verge of AF ordering \cite{KimNatMater2012}.

Our result is compatible with the notion that the low-energy spin excitations in the iron pnictides are affected by itinerant carriers, while the high-energy excitations are primarily TSE arising from localized moments \cite{WangPreprint2013,DaiNPhys2012}. It would be interesting to extend the
spin-polarized measurements to higher energies to determine the evolution of the LSE.
The successful description of our data by the spin-wave theory demonstrates the validity of the local-moment picture for describing the TSE down to the lowest energy.
From previous work \cite{Qureshi_PRB_2012}, we know that
the large in-plane and $c$-axis spin anisotropy disappears above $T_\mathrm{N}$.
In spin-polarized measurements on NaFeAs,
there is evidence for in-plane spin-excitation anisotropy
in the paramagnetic orthorhombic phase \cite{SongUnpublished}, similar to the spin-excitation anisotropy in the tetragonal phase of superconducting BaFe$_{1.904}$Ni$_{0.096}$As$_2$  \cite{LuoPRL2013}.  It would clearly be interesting to determine how the LSE and TSE signals change above $T_\mathrm{N}$ in BaFe$_2$As$_2$.

In summary, we have discovered a LSE $M_a$ signal and determined the TSE $M_b$ and $M_c$ components to a high precision at both the AF zone center and the zone boundary in an iron-pnictide parent compound. Since the $M_a$ component in nearly optimally electron-doped superconductor BaFe$_{1.905}$Ni$_{0.096}$As$_2$ changes dramatically across $T_c$ \cite{LuoPRL2013}, the presence of such a signal in the undoped BaFe$_2$As$_2$ suggests that itinerant electrons, which are important for superconductivity, also contributes to the magnetism in the parent compounds of iron pnictides.

We wish to thank Z.-Y. Weng, D.-X. Yao, and Y.-Z. You for stimulating discussions. Work at Peking University is supported by the National Basic Research Program of China (No. 2013CB921900). Work at IOP, CAS, is supported by the National Basic Research Program of China (Nos. 2011CBA00110 and 2012CB821400) and the National Science Foundation of China (No. 11004233). Work at Rice University is supported by US NSF DMR-1063866.

\bibliography{BFA_reference}


\pagebreak

\section{Supplemental Material}

\setcounter{figure}{0}

\textbf{\textit{Calculation of spin waves.}} The Hamiltonian used here consists of two parts: the Heisenberg Hamiltonian $H_0$ and the anisotropy part $H_1$. $H_0$ has the form:
\begin{equation}
\begin{split}
\label{equation:H0}
&H_0=\sum_{\mathbf{r}}[J_{1a}\mathbf{S}_{\mathbf{r}}\cdot\mathbf{S}_{\mathbf{r}+\hat{x}}+J_{1b}\mathbf{S}_{\mathbf{r}}\cdot\mathbf{S}_{\mathbf{r}+\hat{y}}+
J_{2}(\mathbf{S}_{\mathbf{r}}\cdot\mathbf{S}_{\mathbf{r}+\hat{x}+\hat{y}}+\\
&\mathbf{S}_{\mathbf{r}}\cdot\mathbf{S}_{\mathbf{r}-\hat{x}+\hat{y}})+J_{c}\mathbf{S}_{\mathbf{r}}\cdot\mathbf{S}_{\mathbf{r}+\hat{z}}].
\end{split}
\end{equation}
The contribution to linear spin wave Hamiltonian from $H_0$ is
\begin{equation}
\begin{split}
\label{equation:HLSW}
H_{LSW}=S\cdot\sum_{\mathbf{k} \in BZ}({\hat{b}}^{{\dag}}_{\mathbf{k}},{\hat{b}}_{-\mathbf{k}})
\left( {\begin{array}{*{20}{c}}
   {{A_{\mathbf{k},0}}} & {{B_{\mathbf{k},0}}}  \\
   {{B_{\mathbf{k},0}}} & {{A_{\mathbf{k},0}}}  \\
\end{array}} \right)\binom{{\hat{b}}_{\mathbf{k}}}{{\hat{b}}^{{\dag}}_{-\mathbf{k}}},
\end{split}
\end{equation}
where
\begin{equation}
\begin{split}
\label{equation:A0}
&A_{\mathbf{k},0}=4J_2+2J_{1a}+2J_{1b}(\cos{k_y}-1)+2J_c,\\
&B_{\mathbf{k},0}=4J_2\cos{k_x}\cos{k_y}+2J_{1a}\cos{k_x}+2J_c\cos{k_z}.
\end{split}
\end{equation}

For single-ion anisotropy,
\begin{equation}
\begin{split}
\label{equation:H1ion}
H_1=\sum_{\textbf{r}}[-D_{x}{S_{\textbf{r}}^x}^2-D_{y}{S_{\textbf{r}}^y}^2].
\end{split}
\end{equation}
This term can align the ordered moments to the $\hat{x}$ direction, when $D_{x}>0$ and  $D_{x}>D_y$.
The contribution of $H_1$ to linear spin wave Hamiltonian is similar to  Eq.~(\ref{equation:HLSW}), with $A_{\mathbf{k},0}$, $B_{\mathbf{k},0}$ replaced by the following $A_{\mathbf{k},1}$, $B_{\mathbf{k},1}$:

\begin{equation}
\begin{split}
\label{equation:A1ion}
&A_{\mathbf{k},1}=2D_{x}-D_{y},\\
&B_{\mathbf{k},1}=-D_{y}.
\end{split}
\end{equation}

For magnetic exchange anisotropy,
\begin{equation}
\begin{split}
\label{equation:H1exchange}
H_1=\sum_{\textbf{r}}[D_{1a}S_{\textbf{r}}^{x}S_{\textbf{r}+\hat{x}}^x+D_{1b}S_{\textbf{r}}^{y}S_{\textbf{r}+\hat{y}}^y].
\end{split}
\end{equation}
This term can align the ordered moments to $\hat{x}$ direction when $D_{1a}>0$.
The contribution of $H_1$ to linear spin wave Hamiltonian is similar to Eq.(\ref{equation:HLSW}), with $A_{\mathbf{k},0}$, $B_{\mathbf{k},0}$ replaced by the following  $A_{\mathbf{k},1}$,  $B_{\mathbf{k},1}$:
\begin{equation}
\begin{split}
\label{equation:A1exch}
&A_{\mathbf{k},1}=2D_{1a}+D_{1b}\cos{k_y},\\
&B_{\mathbf{k},1}=D_{1b}\cos{k_y}.
\end{split}
\end{equation}

For both types of anisotropy, the dispersion of spin waves is
\begin{equation}
\begin{split}
\label{equation:spinwave}
E(\mathbf{k})=\sqrt{({A_{\mathbf{k},0}+A_{\mathbf{k},1}})^2-({B_{\mathbf{k},0}+B_{\mathbf{k},1}})^2}
\end{split}
\end{equation}
Based on the above, we obtain the spin-wave dispersion as shown in Table.~\ref{table:two}.

\begin{table*}[htbp]
 \centering
 \begin{tabular}{|c|c|c|c|c|}\hline\hline
~~Response~~       &$(k_x,k_y,k_z)$  &$E(\mathbf{k})/S$  \\\hline
$M_{b,L=1}$	&~~$(\pi,0,\pi)$~~	 &~~$\sqrt{(2D_{x}-2D_{y})(8J_2+4J_{1a}+2D_{x}+4J_c)}$~~	 \\\hline
$M_{c,L=1}$	    &~~$(0,0,0)$~~	    &~~$\sqrt{2D_{x}(8J_2+4J_{1a}+2D_{x}+4J_c-2D_{y})}$~~	 \\\hline
$M_{b,L=2}$    &~~$(\pi,0,0)$~~	 &~~$\sqrt{(2D_{x}-2D_{y}+4J_c)(8J_2+4J_{1a}+2D_{x})}$~~	 \\\hline
$M_{c,L=2}$     &~~$(0,0,\pi)$~~	    &~~$\sqrt{(2D_{x}+4J_c)(8J_2+4J_{1a}+2D_{x}-2D_{y})}$~~	 \\\hline
\hline
 \end{tabular}
\caption{\label{table:two}
Spin-wave dispersion for Hamiltonian with single-ion anisotropy at representative momentum positions and for the two different fluctuation directions. For exchange anisotropy, replace $D_{x}$ by $D_{1a}$, and $D_y$ by $-D_{1b}$.
}
\end{table*}

Using the values of intra-layer $J$ in Ref.\cite{Harriger_PRB_2011} and based on our experimental data, the remaining three parameters can be obtained:
\begin{equation}
\begin{split}
\label{equation:fitexc}
&SD_{1a}=SD_{x}=0.196\,\mathrm{meV},\\
&SD_{1b}=-SD_y=0.311\,\mathrm{meV},\\
&SJ_c=0.333\,\mathrm{meV}.\\
\end{split}
\end{equation}

The two types of anisotropy give identical results at $k_y=0$, which is where our measurements were performed. But even at $k_y=\pi$ where Eqs.~(\ref{equation:A1ion}) and (\ref{equation:A1exch}) are most different, the difference in the spin-wave energies is only about 1 meV in 200 meV, which is nearly impossible to distinguish by inelastic neutron scattering.

\makeatletter
\renewcommand{\thefigure}{S\@arabic\c@figure}

\begin{figure*}[hb]
\includegraphics[width=5in]{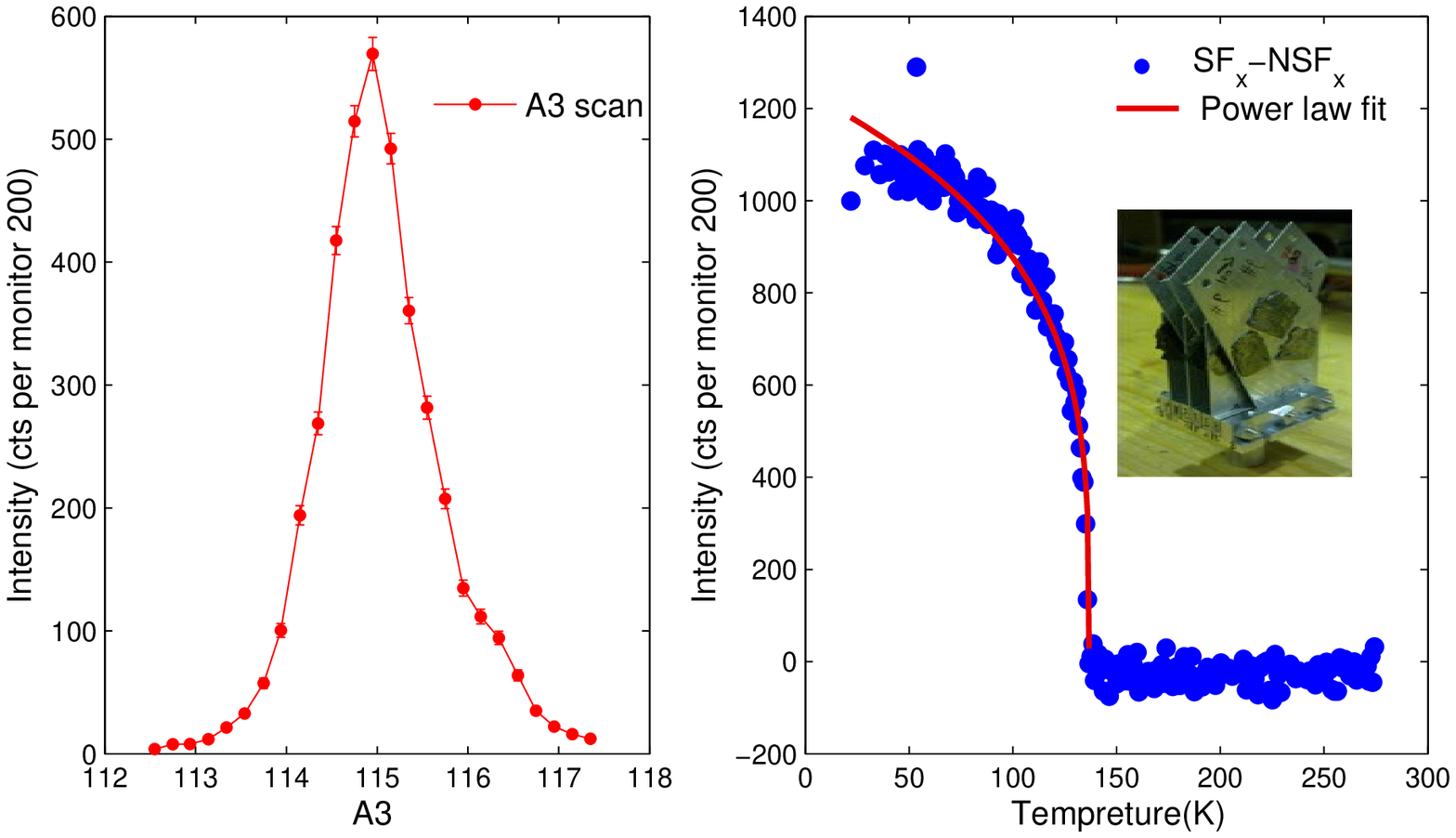}
\caption{\label{sfig:two}
\textbf{\textit{Sample information.}} (a) Rocking scan through a nuclear Bragg peak measured with $k_f=2.662\mathrm{\AA}^{-1}$. The sample mosaic is $1.2^\circ$ as indicated by the full width at half maximum of the peak. (b) Magnetic Bragg peak intensity measured at $(1~0~3)$ in the spin-flip and non-spin-flip geometries with changing temperature. Fitting the intensity difference between the two geometries to a power law yields a spin-density-wave temperature of 137 K. Inset is a photo of the sample.
}
\end{figure*}

\begin{figure*}[hb]
\includegraphics[width=3.375in]{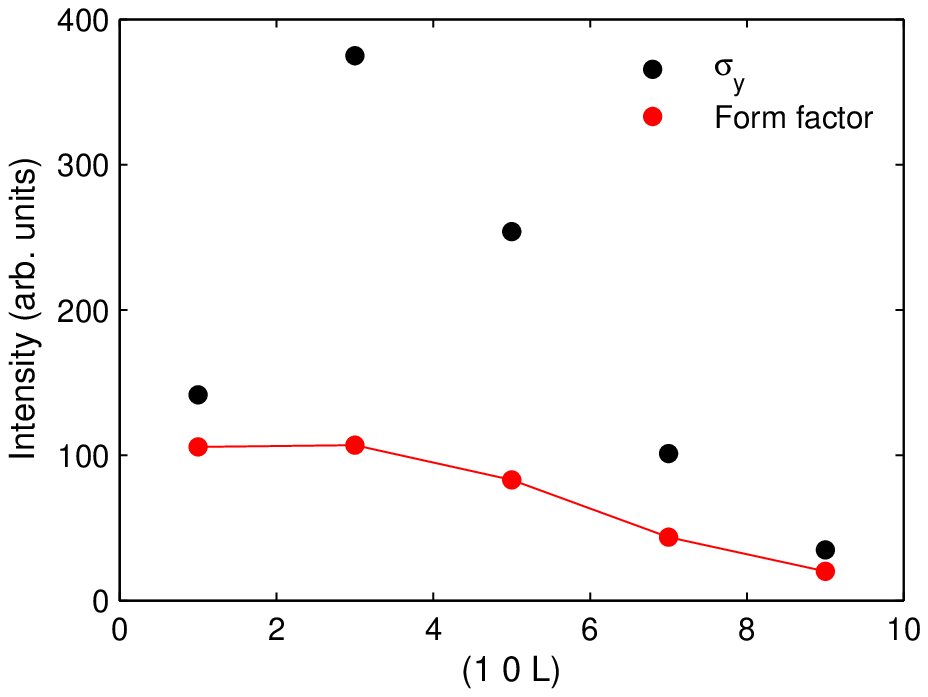}
\caption{\label{sfig:four}
\textbf{\textit{Determination of the magnetic form factor.}} Magnetic Bragg peak intensities measured at $(1~0~L)$ with $L=1$, 3, 5, 7, and 9. The plotted $\sigma_y$ (black circles) is deduced from raw data of all six (spin-flip and non-spin-flip, $\hat{x}$, $\hat{y}$, and $\hat{z}$) polarization geometries. The form factor (red circles) is calculated from $\sigma_y$ after corrections for the orientation factor and the resolution factor (the so-called $R_0$). It is seen that the form factor does not change much from $L=1$ to $L=3$, consistent with our inelastic measurements.
}
\end{figure*}

\begin{figure*}[!htb]
\includegraphics[width=6.7in]{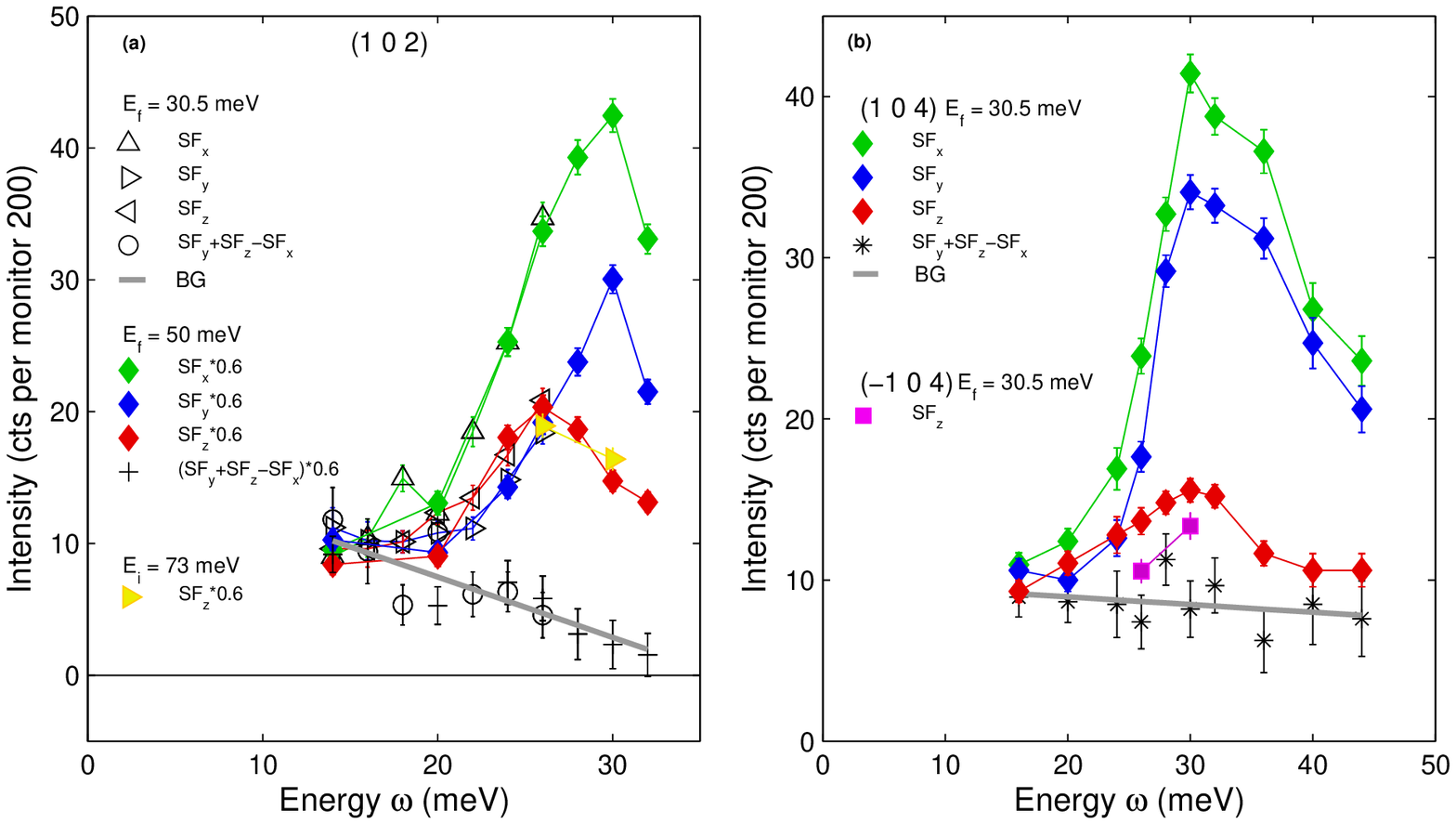}
\caption{\label{sfig:one}
\textbf{\textit{Raw data obtained for the AF zone boundaries and test of resolution effects.}} Raw SF data obtained at the AF zone boundary $L=2$ (a) and $L=4$ (b), from which the magnetic signals in Fig.~3 of the main text are extracted. It can be seen that the data measured with $E_\mathrm{f}=30.5$ and 50 meV agree reasonably well after normalization (the 50 meV data are multiplied by 0.6 which accounts for the change in the resolution volume). To further verify that the maxima of the SF$_z$ data are at 26 and 30 meV for $L=2$ and 4, respectively, additional measurements were performed in order to rule out a distortion of the data due to resolution effects: For $L=2$, measurements with fixed incident-neutron energy $E_\mathrm{i}=73$ meV were performed at 26 and 30 meV (yellow triangles), which confirm that the intensity is higher at 26 meV, and that the decrease of intensity above 26 meV is not because of a resolution change due to the changing $E_\mathrm{i}$ in the $E_\mathrm{f}=50$ meV measurement. For $L=4$, measurements at 26 and 30 meV (magenta squares) confirm that the intensity is higher at 30 meV for both $(1~0~4)$ and $(-1~0~4)$, which rules out resolution focusing artefacts. These tests show no evidence for a distortion of data by resolution effects. Measurements with fixed $E_\mathrm{i}$ and with the chosen fixed $E_\mathrm{f}$ are possible because the neutron guide removes all high-order incident neutrons with energy greater than $\sim120$ meV, and no PG filter is required in most of our measurements.
}

\end{figure*}

\begin{figure*}[!htb]
\includegraphics[width=3.375in]{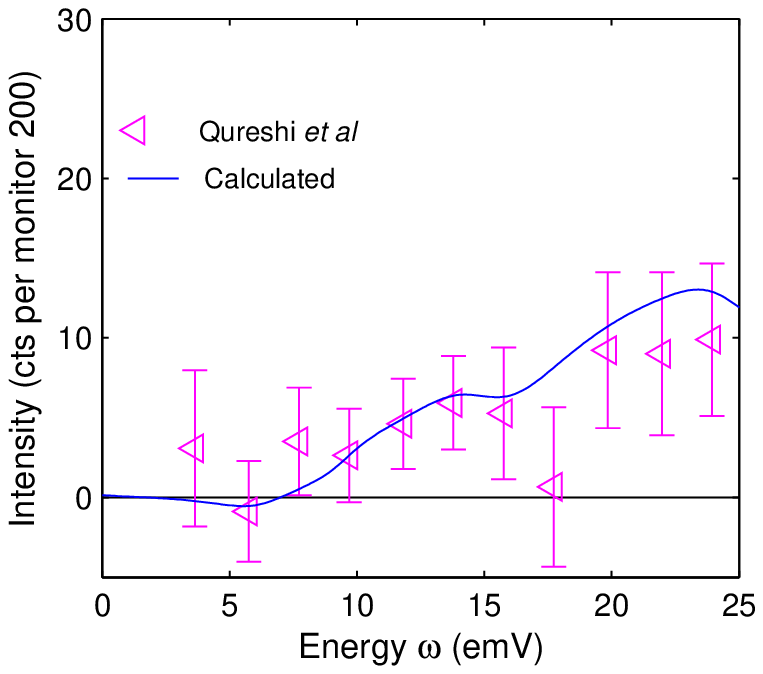}
\caption{\label{sfig:three}
\textbf{\textit{Comparison to an earlier report.}} $\sigma_y$ at $(1~0~5)$ calculated from the $M_c$ and $M_a$ values we obtained (blue line) compared to data reported by Qureshi \textit{et al.} \cite{Qureshi_PRB_2012} (data points) after normalization according to the $\sigma_z$ values of the two experiments. In spite of the large statistical uncertainty, the previously reported data agree reasonably well with the ``prediction'' based on our result.
}
\end{figure*}

\end{document}